\documentclass[12pt]{article}
\usepackage{epsfig}
\usepackage{cite}

\newcommand{\beq}{\begin{equation}}
\newcommand{\eeq}{\end{equation}}
\newcommand{\bea}{\begin{eqnarray}}
\newcommand{\eea}{\end{eqnarray}}

\def\fig#1{Fig.~{\ref{#1}}}

\newcommand\sss{\scriptscriptstyle}
\newcommand\as{\alpha_{\sss S}} 
\newcommand\aem{\alpha_{\rm em}}

\def\epem{e^+e^-}
\def\ord{{\cal O} }
\def\MS{$\overline{\rm MS}$}
\def\bentarrow{\:\raisebox{1.3ex}{\rlap{$\vert$}}\!\longrightarrow}

  \newcommand{\ccaption}[2]{
    \begin{center}
    \parbox{0.85\textwidth}{
      \caption[#1]{\small{\it{#2}}}
      }
    \end{center}
    }
\begin{document}

\begin{titlepage}

\vspace{2cm}
\begin{center}
{\Large\bf Physics at Large Rapidities in $\gamma^*\gamma^* \to$ 
Hadrons at LEP2}\\
\vspace{2.cm}

{\large
{M. Cacciari$^1$, V. Del Duca$^2$\footnote{Rapporteur at the LEPTRE
meeting, Univ. di Roma Tre, Roma 2001.}, S. Frixione$^3$\footnote{On leave
of absence from I.N.F.N., Sezione di Genova, Italy} and
Z. Tr\'ocs\'anyi$^4$\footnote{Sz\'echenyi fellow of the Hungarian
Ministry of Education}}
}

\vspace{.2cm}
{$^1$\sl Dipartimento di Fisica, Universita' di Parma\\ and
INFN, Sezione di Milano, Gruppo Collegato di Parma}\\

\vspace{.2cm}
{$^2$\sl I.N.F.N., Sezione di Torino\\
via P. Giuria, 1\ \ 10125 Torino, Italy}\\

\vspace{.2cm}
{$^3$\sl LAPTH, BP110\\ F-74941, Annecy le Vieux Cedex, France}\\

\vspace{.2cm}
{$^4$\sl University of Debrecen and\\
Institute of Nuclear Research of the Hungarian Academy of Sciences\\ 
H-4001 Debrecen, PO Box 51, Hungary}\\

\vspace{.5cm}

\begin{abstract}
We compute the order-$\as$ corrections to the total 
cross section and to jet rates in $\gamma^* \gamma^* \to$~hadrons
for the process $\epem\to\epem +$~hadrons.
We use a NLO order general-purpose partonic Monte Carlo 
event generator that allows the computation of a rate differential 
in the produced leptons and hadrons. We compare our results with the 
experimental data for $\epem\to\epem +$~hadrons at LEP2.
\end{abstract}

\end{center}
 \vfil

\end{titlepage}

Strong interaction processes, characterised by a large kinematic scale,
are described in perturbative QCD by a fixed-order expansion in $\as$
of the parton cross section, complemented, if the scattering process is
initiated by strong interacting partons, with the Altarelli-Parisi evolution
of the parton densities. In many scattering processes, the-state-of-the-art
computation of production rates is at the next-to-leading order (NLO). 
However, in kinematic regions characterised by two large and disparate 
scales, a fixed-order expansion may not suffice: large logarithms of the
ratio of the kinematic scales appear, which may have to be resummed.
In processes where the centre-of-mass energy $S$ is much larger than the
typical transverse scale $Q^2$, the leading logarithms of type $\ln(S/Q^2)$
are resummed by the BFKL equation.
Several observables, like the scaling
violations of the $F_2$ structure function,
forward-jet production in DIS,
dijet production at large rapidity intervals
and $\gamma^* \gamma^* \to$ hadrons in $\epem$ 
collisions~\cite{Acciarri:1999ix,opal,Bartels:1996ke,
Cacciari:2001cb} have been measured and analysed in this fashion.

In this talk we report on a NLO calculation~\cite{Cacciari:2001cb} of the 
total cross section and
of jet rates in $\gamma^* \gamma^* \to$ hadrons for the process
$\epem\to\epem +$~hadrons at LEP2, and we compare the NLO calculation to 
the data from the CERN L3~\cite{Acciarri:1999ix} and 
OPAL~\cite{opal} collaborations. Namely, we consider
\beq
\begin{array}{rcl}
e^+ + e^- & \longrightarrow & e^+ + e^- + \underbrace{\gamma^* + \gamma^*} \\
 &  & \phantom{e^+ + e^- + \gamma^*\:}\bentarrow {\rm hadrons} ;
\end{array}
\label{processee}
\eeq
selecting those events in which the incoming leptons produce two 
photons which eventually initiate the hard 
scattering that produces the hadrons. However, it is clear that the 
process in Eq.~(\ref{processee}) is non physical; rather, it has to be 
understood as a shorthand notation for a subset of Feynman diagrams 
contributing to the process that is actually observed,
\beq
e^+ +e^-\,\longrightarrow\,
e^+ +e^- + {\rm hadrons}.
\label{fullproc}
\eeq
Other contributions to the process in Eq.~(\ref{fullproc}) are, for example, 
those in which the incoming $\epem$ pair annihilates into a photon or a $Z$
boson, eventually producing the hadrons and a lepton pair, or those in which 
one (or both) of the two photons in Eq.~(\ref{processee}) is replaced
by a $Z$ boson. However, one can devise a set of cuts such
that the process in Eq.~(\ref{processee}) gives the only non-negligible
contribution to the process in Eq.~(\ref{fullproc}). One can tag 
both of the outgoing leptons, and retain only those events in which the 
scattering angles 
of the leptons are small: in such a way, the contamination due 
to annihilation processes is safely negligible.
Furthermore, small-angle tagging also guarantees that the photon
virtualities are never too large (at LEP2, one typically measures
$Q_i^2={\cal O}(10$~GeV$^2$)); therefore, the contributions from processes
in which a photon is replaced by a $Z$ boson are also negligible.
Thus, it is not difficult to extract the cross section of the 
process \mbox{$\gamma^*\gamma^* \to$~hadrons} from the data relevant 
to the process in Eq.~(\ref{fullproc}).

Our calculation was performed in the massless limit for the final state
quarks. We compared our LO result
to the massless limit of the JAMVG program of Ref.~\cite{Vermaseren:1983cz},
and found perfect agreement.
To study the effect of the NLO corrections, we used the experimental cuts
employed by the L3 Collaboration. The scattered 
electron and positron are required to have energy $E_{1,2}$ larger than 30 
GeV and scattering angle $\theta_{1,2}$ between 30 and 66 mrad. Furthermore, 
the rapidity-like variable $Y$, defined by
\beq
Y=\log\frac{y_1 y_2 S}{\sqrt{Q_1^2 Q_2^2}} ,
\label{YD}
\eeq
is required to lie between 2 and 7 ($y_i$, with $i$~=1, 2, is proportional 
to the
light-cone momentum fraction of the $i^{th}$ virtual photon, and is precisely
defined in Ref.~\cite{Cacciari:2001cb}, 
where a discussion on the properties of $Y$ can also be found). The cross 
sections have been evaluated at $\sqrt{S} = 200$ GeV, including up to
five massless flavours.

\begin{figure}[t]
\begin{center}
\epsfig{file=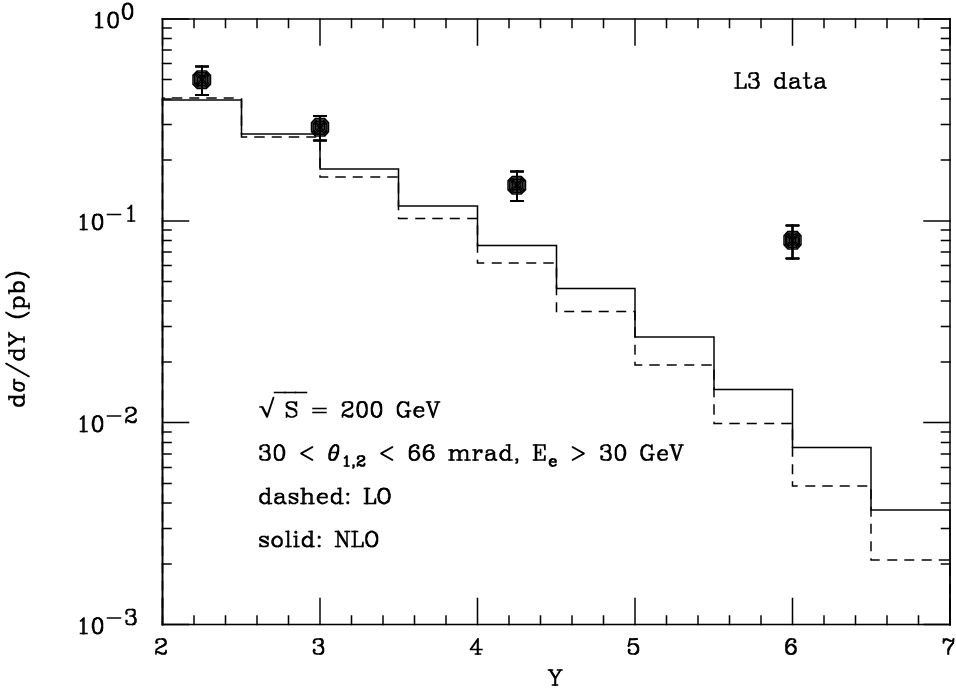,width=11cm}
\ccaption{}{\label{fig-l3}
Differential cross section with respect to $Y$ from the L3
Collaboration compared to leading and next-to-leading order
predictions. The data are taken at $\sqrt{S} = 189 - 202$~GeV. 
The theoretical simulation is always run at $\sqrt{S} = 200$ GeV.}
\end{center}
\end{figure}

We discuss briefly the dependence of our predictions on the
electromagnetic and strong couplings;
our cross sections are $\ord(\aem^4)$ and
we chose to evolve $\aem$ (using one-loop \MS~running)
on an event-by-event basis to the scales set by the virtualities of
the exchanged photons; hence, we replace the Thomson value $\alpha_0
\simeq 1/137$ by $\aem(Q_i^2)$. We treat independently the two photon 
legs: thus, $\alpha_{\rm em}^4$ has to be 
understood as \mbox{$\alpha_{\rm em}^2(Q_1^2)\alpha_{\rm em}^2(Q_2^2)$}.
As for the strong coupling $\as$, we define a default scale $\mu_0$ so 
as to match the order of magnitude of the (inverse of the) interaction 
range,
\beq
\mu_0^2 = \frac{Q_1^2+Q_2^2}{2} +
\left(\frac{k_{1{\sss T}} + k_{2{\sss T}} + k_{3{\sss T}}}{2}\right)^2 \: .
\label{defscale}
\eeq
The renormalization scale $\mu$ entering $\as$ is set 
equal to $\mu_0$ as a default value, and equal to $\mu_0/2$ or $2\mu_0$ when
studying the scale dependence of the cross section. In Eq.~(\ref{defscale}),
the $k_{i{\sss T}}$ are the transverse energies of the outgoing quarks and, 
for three-particle events, the emitted gluon. Since the hard process is 
initiated by the two virtual photons, we study its
properties in the $\gamma^*\gamma^*$ center-of-mass frame. 
We evolve $\as$ to next-to-leading log accuracy, 
with $\as(M_{\sss Z})=0.1181$~\cite{Groom:2000in} (in $\overline{{\rm MS}}$ 
at two loops and with five flavours, this implies 
$\Lambda_{\overline{{\rm MS}}}^{(5)} = 0.2275$ GeV).
In all of the distributions examined~\cite{Cacciari:2001cb}, we found that
the uncertainty related to $\mu$ is always smaller than 
the net effect of including the NLO corrections.
As for the effect of the NLO corrections themselves, we found that, apart
from slightly increasing the cross section, they induce visible shape
modifications in the $Y$ distribution (see \fig{fig-l3}):
their effect changes from almost nil at the left edge of
the plots to a more than 50\% increase at the right one.

%%%%%%%%%%%%%%%%%%%%%%%%%%%%%%%%%%%%%%%%%%%%%%%%%%%%%%%%%%%%%%%%%%%%%%
\begin{table}[htb]
\begin{center}
\begin{tabular}{|c|c|c|c|}
\hline
\multicolumn{4}{|c|}{$d\sigma/dY$ (pb)~~~~~~~ $\sqrt{S} = 189 - 202$
GeV}\\
\hline
%$\Delta Y$&L3 Data                     & LO    & NLO  & NLO mass corr. \\
%\hline
%2.0 -- 2.5& 0.50 $\pm$ 0.07 $\pm$ 0.03 & 0.405  & 0.395 & 0.356\\
%2.5 -- 3.5& 0.29 $\pm$ 0.03 $\pm$ 0.02 & 0.212  & 0.224 & 0.204\\
%3.5 -- 5.0& 0.15 $\pm$ 0.02 $\pm$ 0.01 & 0.067  & 0.0800& 0.0704\\
%5.0 -- 7.0& 0.08 $\pm$ 0.01 $\pm$ 0.01 & 0.0090 & 0.0131& 0.0109\\
$\Delta Y$&L3 Data                     & LO    & NLO  \\
\hline
2.0 -- 2.5& 0.50 $\pm$ 0.07 $\pm$ 0.03 & 0.405 &0.396$^{+0.002}_{-0.002}$ \\
2.5 -- 3.5& 0.29 $\pm$ 0.03 $\pm$ 0.02 & 0.213 &0.225$^{+0.001}_{-0.002}$ \\
3.5 -- 5.0& 0.15 $\pm$ 0.02 $\pm$ 0.01 & 0.067&0.080$^{+0.002}_{-0.002}$\\
5.0 -- 7.0& 0.08 $\pm$ 0.01 $\pm$ 0.01 & 0.0091&0.0131$^{+0.0009}_{-0.0006}$\\
\hline
\hline
Total & 0.93 $\pm$ 0.05 $\pm$ 0.07  & 0.534  & 0.569$^{+0.006}_{-0.004}$ \\
\hline
\end{tabular}
\ccaption{}{\label{table2} 
The experimental cross section from L3 compared to leading and 
next-to-leading order predictions. The uncertainties in the NLO column
are related to variations of the renormalization scale.
}
\end{center}
\end{table}
%%%%%%%%%%%%%%%%%%%%%%%%%%%%%%%%%

The L3~\cite{Acciarri:1999ix} and OPAL~\cite{opal} collaborations have
analysed data for hadron production in $\epem$ 
collisions (through $\gamma^*\gamma^*$ scattering) at a center-of-mass 
energy around 200 GeV. L3 made use of the previously mentioned set of experimental cuts. The
cross section they find, as a function of $Y$, is reported in
Table~\ref{table2} and plotted in Fig.~\ref{fig-l3}. Table~\ref{table2} 
shows, in four different $Y$ bins, the experimental cross section compared
to our leading and next-to-leading order predictions, evaluated at $\sqrt{S}
= 200$~GeV. The same comparison is made in Fig.~\ref{fig-l3}: the data 
lie above the theory in the low-$Y$ region, and sizably overshoot the 
predictions in the large-$Y$ one. Thus we find a marked difference in 
shape between theory and data which, if confirmed, could be interpreted 
as the onset of important higher order effects, perhaps of BFKL type.
As can be seen from Table~\ref{table2}, the scale uncertainties
affecting our predictions are much smaller than the experimental
errors. 
We have also studied the effect of the finite mass of the outgoing
heavy quarks in the charm and bottom case, by comparing our results
with the ones obtained with the JAMVG~\cite{Vermaseren:1983cz} code.
Within the L3 set of cuts, such mass effects can be seen to decrease
the LO massless cross section by an amount of the order of  10-15\%.

In Ref.~\cite{Cacciari:2001cb} we have also compared our prediction to 
the data~\cite{opal} the OPAL Collaboration has 
taken at $\sqrt{S} = 189$ - 202 GeV, making use of
a slightly  different set of cuts. 
The prediction for the total cross section falls short of the
central OPAL result, but is well within the experimental error.
For the differential distribution in the variable
$\overline{Y}$ (a variant of $Y$, which tends to it in the high-energy
limit~\cite{Cacciari:2001cb}), a generally good 
agreement within errors can be observed, even though in the largest
$\overline{Y}$ bin the data tend to lie above the prediction.
Given the large discrepancy
between theory and L3 data for the distribution in $Y$, it shall therefore
be of utmost importance to measure as accurately as possible the $Y$ spectrum,
in order to perform a precise study of effects beyond NLO (such as BFKL 
dynamics).

%%%%%%%%%%%%%%%%%%%%%%%%%%%%%%%%%%%%%%%%%%%%%%%%%%%%%%%%%%%

%%%%%%%%%%%%%%%%%%%%%%%%%%%%%%%%%%%%%%%%%%%%%%%%%%%%%%%%
% End of the paper
%
\end{document}